\documentclass[12pt,preprint]{aastex}

\slugcomment{Revised: ApJ, March 30, 2011}

\shorttitle{Cepheid Sensitivity to Metallicity}
\shortauthors{Freedman \& Madore}

\begin{document}


\title{Two New Tests of the Metallicity Sensitivity of the
\\ Cepheid Period-Luminosity Relation (The Leavitt Law)}


\author{\bf   Wendy L. Freedman \& Barry F. Madore}
\affil{Observatories of the Carnegie Institution of Washington \\ 813
Santa Barbara St., Pasadena, CA ~~91101} \email{wendy@obs.carnegiescience.edu,
barry@obs.carnegiescience.edu}



\begin{abstract}
We undertake a new test of the metallicity sensitivity
of the Leavitt Law for Classical Cepheids.  We derive an empirical
calibration of the apparent luminosities of Cepheids as measured from
the optical through the mid-infrared (0.45 -- 8.0$\mu$m) as a function
of spectroscopic [Fe/H] abundances of individual Cepheids in the Large
Magellanic Cloud from Romaniello et al. (2008). The cumulative trend
over the entire wavelength range shows a nearly monotonic
behavior. The sense of the trend is consistent with differential
line-blanketing in the optical, leading to stars of high metallicity
being fainter in the optical.  This is followed by a reversal in the
trend at longer wavelengths, with the cross-over occurring near the K
band at about 2.2$\mu$m, consistent with a subsequent redistribution
of energy resulting in a mild brightening of Cepheids (with increased
metallicity) at mid-infrared wavelengths. This conclusion agrees with
that of Romaniello et al. based on a differential comparison of the
mean V- and K-band Leavitt Laws for the Galaxy, SMC and LMC, but is
opposite in sign to most other empirical tests of the sensitivity of
Cepheid distances to mean [O/H] HII region abundances.

We also search for a correlation of Cepheid host-galaxy metallicity with
deviations of the galaxy's Cepheid distance from that predicted from a
pure Hubble flow.  Based on Cepheid distances to 26 nearby galaxies in
the local flow, only a very weak signal is detected giving
$\delta\mu_o = -0.17 (\pm 0.31) ([O/H] - 8.80) - 0.21 (\pm
0.10)$. This is in agreement with previous determinations, but
statistically inconclusive.

\end{abstract}


\vfill\eject
\section{Introduction}

The Cepheid period-luminosity relation (Leavitt Law) remains central
in the determination of extragalactic distances and determination of
the Hubble constant. An outstanding remaining issue in the calibration
of the Cepheid distance scale is the effect of metallicity on Cepheid
luminosities and colors.  Early modeling efforts dating at least as
far back as Robertson (1973) and Iben \& Tuggle (1975) have investigated
the role of metals in the atmospheres of Cepheid variables. More
recent and extensive linear and non-linear pulsation models by a
number of groups have come to differing conclusions on both the sign
and the magnitude of a metallicity effect (see Bono et al. 2008 and
references therein for a recent discussion); hence empirical studies
remain needed.

Over the past two decades we have proposed a number of tests for and
calibrations of this effect on the Cepheid Period-luminosity relation;
hereafter referred to as the Leavitt Law. The first test (Freedman \&
Madore 1990) involved using Cepheids radially distributed across a
single galaxy with a measured chemical composition gradient across
its disk. By examining the change in the (reddening-corrected)
distance moduli of individual Cepheids or groups of Cepheids as a
function of galactocentric radius one can test for the possible
effects due to metallicity. The original test was applied to M31 where no
statistically signficant evidence for a metallicity effect was detected. A
subsequent calibration was undertaken as part of the HST Key Project
(Freedman 2001) using Cepheids in M101 (Kennicutt et al. 1998) this
time giving $\delta (m-M)_0/\delta[O/H]$ = -0.24 $\pm$ 0.16
mag/dex. Another, more recent application of this technique can be
found in Macri et al. (2006) in their study of radially distributed
Cepheids in the nearby galaxy NGC~4258, where they find a slope of
-0.29 $\pm$ 0.09 mag/dex. And finally, Scowcroft et al. (2009) have
examined Cepheids in four fields in the Local Group galaxy M33, and
determined a metallicity sensitivity of -0.29 $\pm$ 0.11 mag/dex.

The second test (Lee, Freedman \& Madore 1993) proposed that a
comparison of the tip of the red giant branch distances (Population
II) with the Cepheid (Population I) distances could be used to look
for a trend in the differential distance moduli as a function of the
host galaxy (HII region) metallicity. This test was successfully
deployed most recently by Sakai et al. (2004) where the slope of the
metallicity relation was determined to be -0.24 $\pm$ 0.05 mag/dex.

The conclusion from these two types of tests is that there is a mild
sensitivity of the zero point of the Leavitt Law to metallicity with
published values broadly centered around -0.25 mag/dex, trending in
the direction that more metal-rich Cepheids are intrinsically brighter
than their metal-poor counterparts (at optical wavelengths). That is,
using a low-metallicity calibration of the Leavitt Law to estimate
distances to high-metallicity Cepheids would result in an
underestimate of their true distances. 

In a different approach, Romaniello et al (2008) have obtained direct
spectroscopic [Fe/H] abundances for a sample of Galactic, LMC and SMC
Cepheids. They compare the Leavitt Law for samples of stars with
different mean metallicities. In contrast to the studies above, they
find that metal-rich Cepheids in the V-band are fainter than
metal-poor ones.

In the following we propose a further test using the published
Romaniello et al (2008) spectroscopic measurements of [Fe/H] to
examine a correlation of the residuals in multi-wavelength photometry
for individual Cepheids in the Large Magellanic Cloud. We also test
the prediction that a metallicity effect will impose correlated
scatter to the observed Hubble diagram.

\section{Two New Tests}
\subsection{The LMC Spectroscopic Sample}

Recently, Romaniello et al. (2008) published iron-line metallicities
for 32 Galactic, 22 Large and 14 Small Magellanic Cloud Cepheids.  They
looked for a relation between the [Fe/H] abundance and the mean V-band
and K-band residuals from the Freedman et al. (2001) and Persson et
al. (2004) Leavitt Laws, respectively. In contrast with previous
studies, they found an increasing dependence of the V-band residuals
with [Fe/H] abundance; while at K, the existence of an effect was less
clear.

In this paper, we explore the run of multi-wavelength (UBVJHK and 3.6, 4.5, 5.8
and 8.0 $\mu $m) residuals from the Leavitt Law for the 22 individual
LMC Cepheids\footnote{The SMC sample was not considered because of the
large additional scatter imposed on all of the SMC Cepheid PL
relations due to back-to-front geometric effects (e.g., Welch,
McLaren, Madore \& McAlary 1987). } with measured [Fe/H]
abundances derived by Romaniello et al. (2008) from
high-resolution (R=30,000) VLT UVES spectra.

\subsubsection{Multi-Wavelength Solutions}

Before searching for a potential sensitivity of Cepheid magnitudes to
metallicity, we first test for a correlation of metallicity with
period. A correlation of this type could be sample-induced; that is
accidental, because of small-number statistics. Or, there could
plausibly be an age effect, where metallicity increases with time,
since periods are tracers of age for Cepheids (e.g., Efremov 1978,
Magnier et al. 1997, Bono et al. 2005). In any event, as Figure 1
shows, neither of these effects appear to be significant; there is no
obvious correlation of metallicity with period.

\begin{figure}
\includegraphics [width=8cm, angle=270] {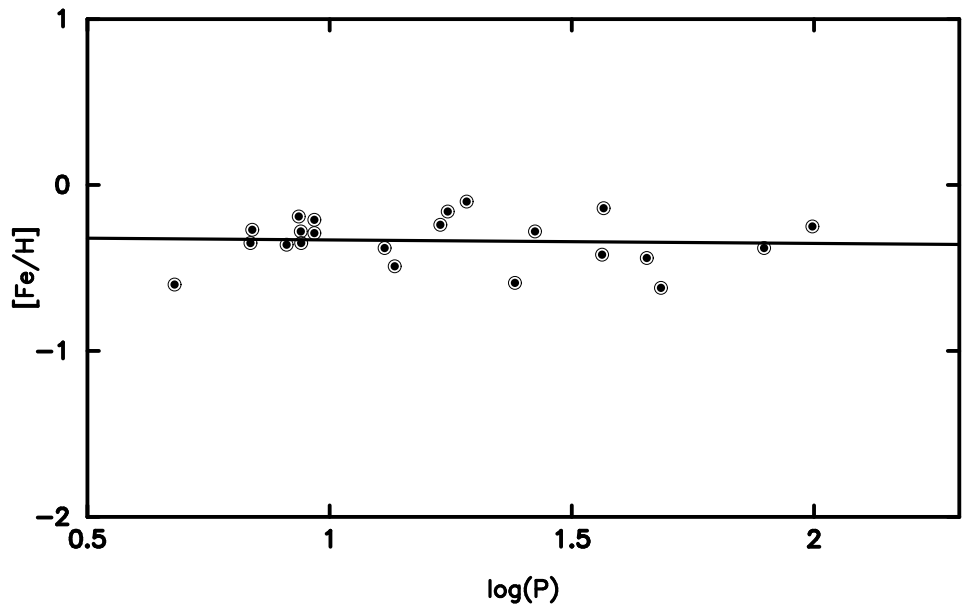}
\caption{A plot of spectroscopic [Fe/H] metallicities as a function of
period for 22 LMC Cepheids observed by Romaniello et al. (2008). No
trend is seen in the data.}
\end{figure}

\begin{figure}
\includegraphics [width=18cm, angle=270] {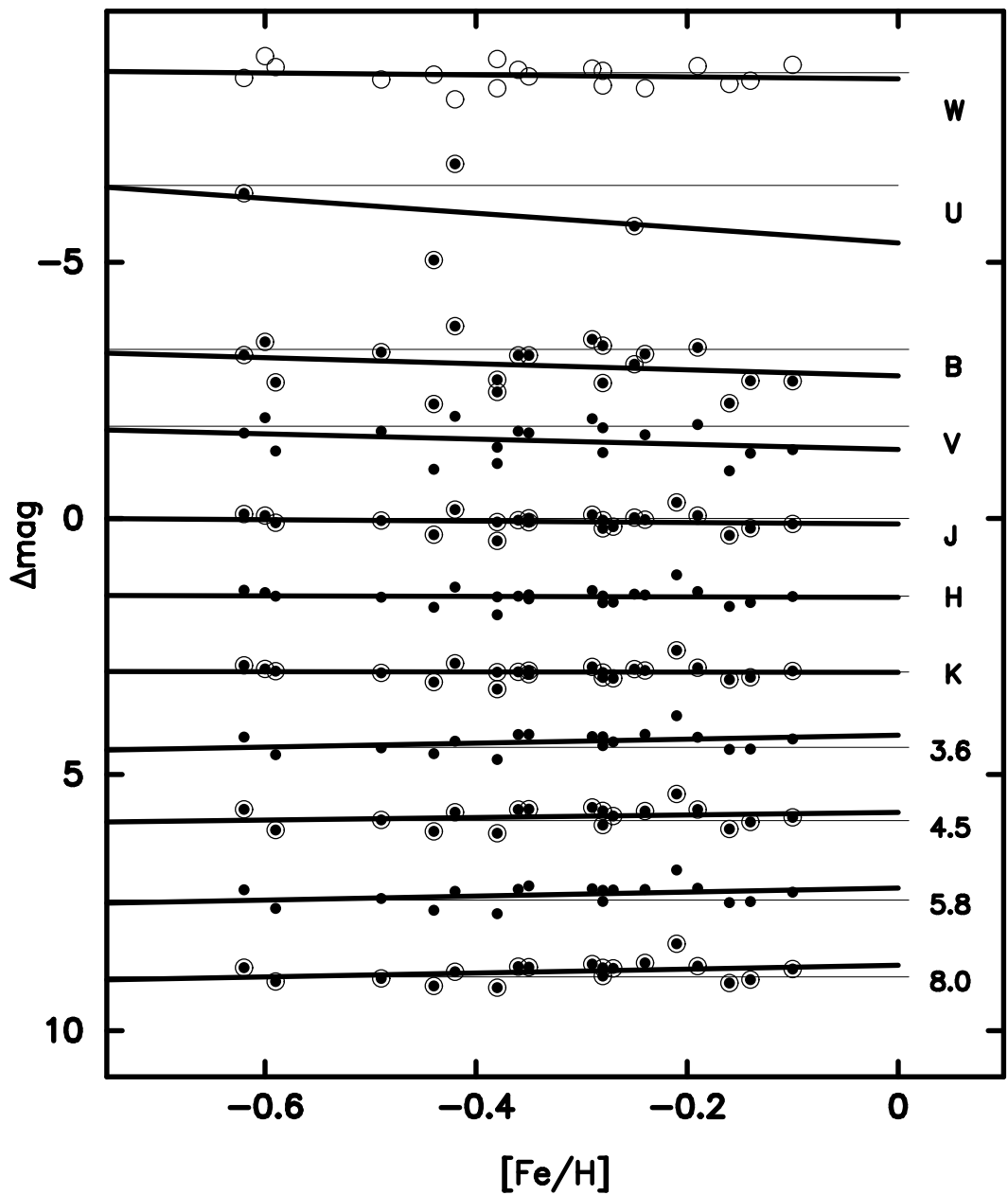}
\caption{A plot of (UBVJHK and 3.6, 4.5, 5.8 and 8 $\mu $m)PL relation
magnitude residuals for individual Cepheids as a function of each
star's individually determined spectroscopic [Fe/H] metallicity. At
the top of the plot, residuals for the reddening-free magnitude W = V-
$R_V$(B -V) are also shown. Linear fits to these residuals are shown
by the thick solid lines. A thin horizontal line is shown for
comparison in each case. Note the strong correlation from bandpass to
bandpass of deviations of individual stars around each of the
regressions.}
\end{figure}

\begin{figure}
\includegraphics [width=18cm, angle=270] {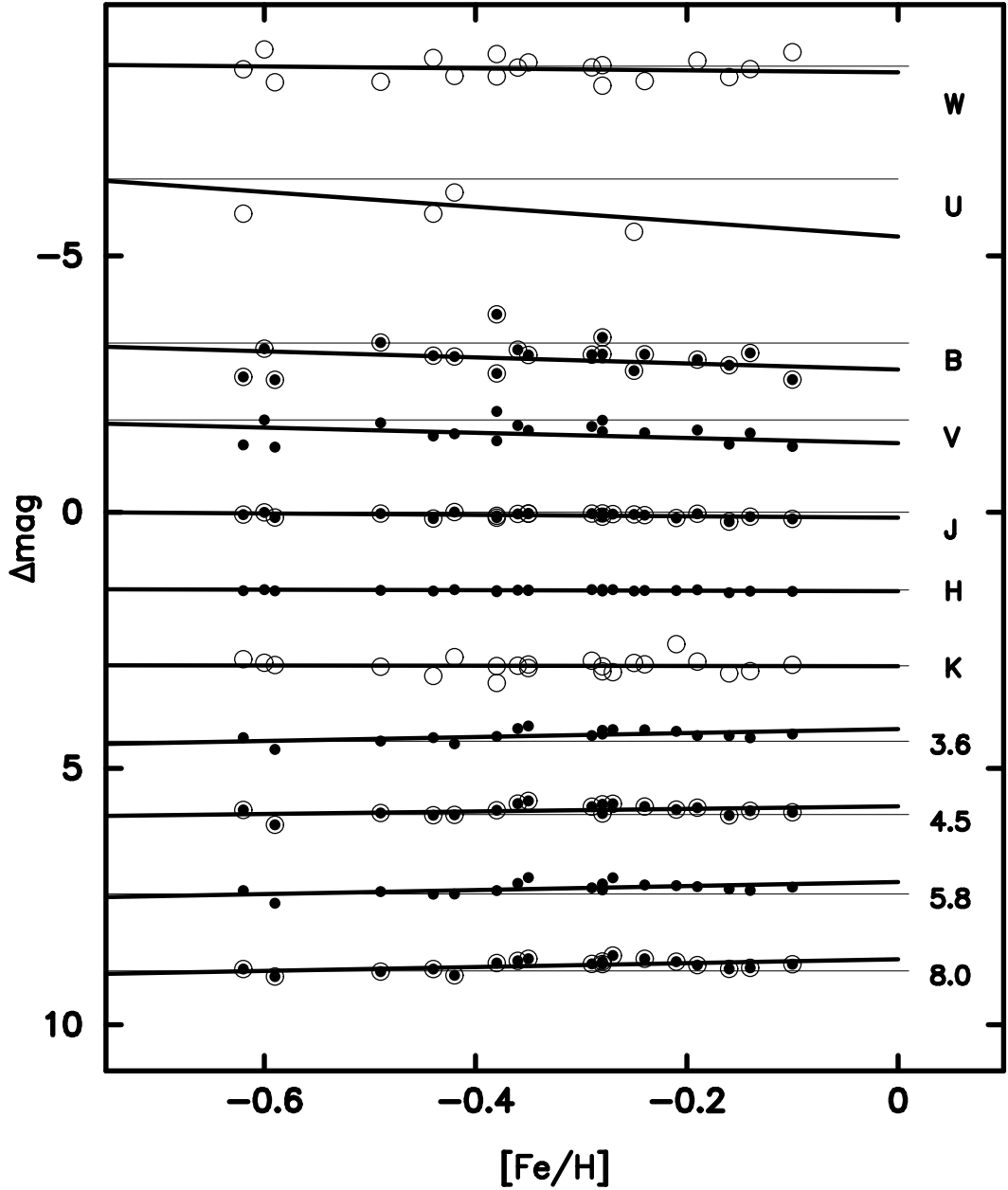}
\caption{The same data as in Fig. 2 but now with the residuals
decorrelated using the H-band data as fiducial. The slopes are
preserved from the fits to the data in Figure 2. The significance of
these slopes is now greatly enhanced.}
\end{figure}

\begin{figure}
\includegraphics [width=8cm, angle=270] {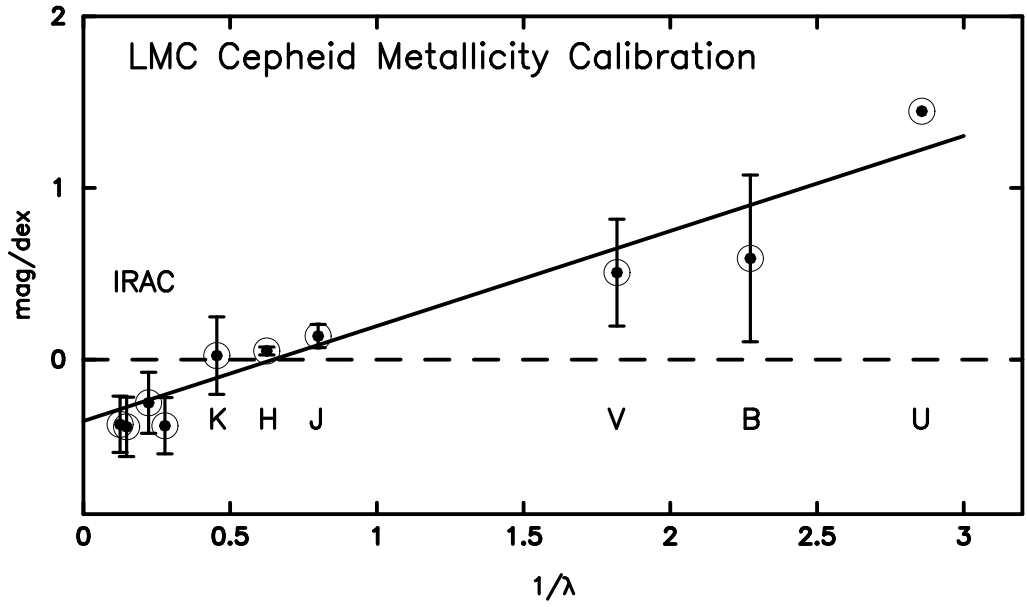}
\caption{The sensitivity of Cepheid magnitudes to metallicity as a
function of wavelength. The slopes derived from the plots in Figure 1
are shown as a function of bandpass (expressed as the inverse
wavelength).  The error bars are from the noise-decorrelated data. The
line is a unweighted fit, (excluding the low-significance U-band data
point to the far right), designed simply to emphasize the trend}.
\end{figure}

\begin{figure}
\includegraphics[width=9cm, angle=-90]{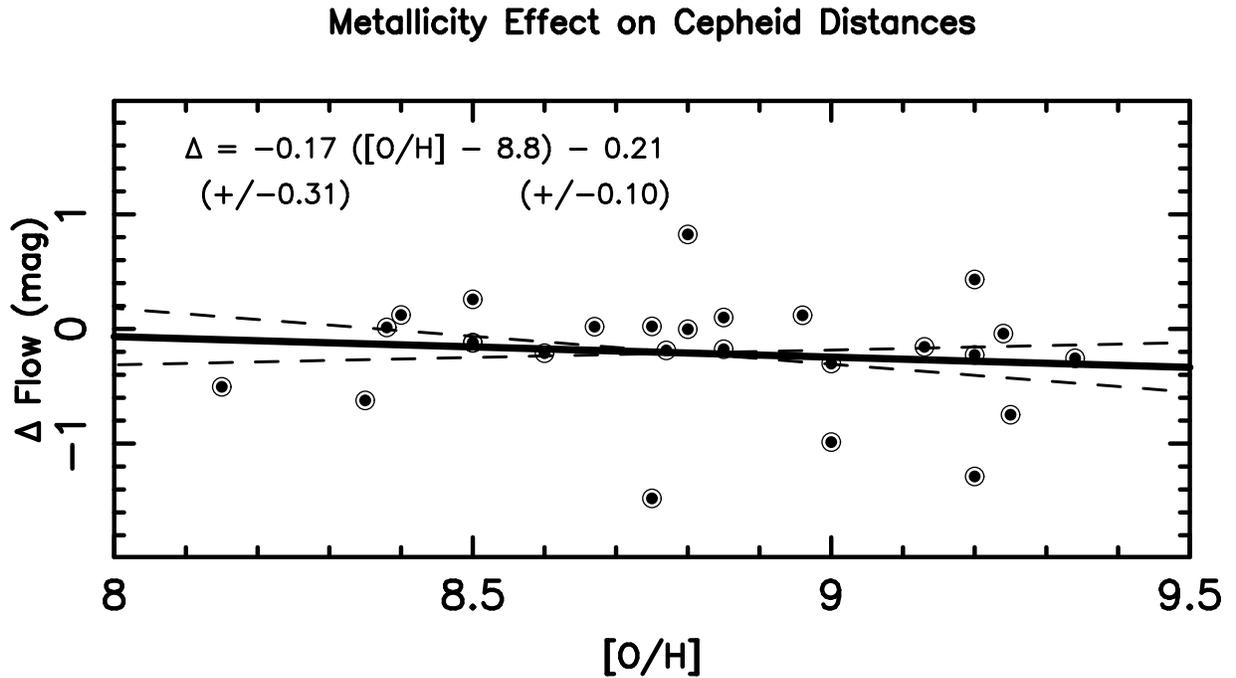}
\caption{Distance modulus deviations of individual galaxies around a
pure Hubble flow, corrected for bulk motions induced by Virgo, the
Great Attractor and the Shapley Constellation, plotted as a function
of the host galaxy metallicity. The sense of the residuals is that
positive residuals indicate that measured distances are greater than
distances predicted for a quiet flow. The trend with metallicity is
such that high-metallicity Cepheids are measured to be too close when
using a low-metallicity PL relation.} 
\end{figure}

We now undertake a test for metallicity, enabled for the first time by
this new spectroscopic dataset, in combination with existing
multi-wavelength photometry. The $UBV$ photometry for the Cepheids is
from Madore (1985); the $JHK$ photometry is from Persson et
al. (2004), and the mid-infrared (3.6 to 8.0$\mu$m) photometry is from
Madore et al. (2009).  For each individual Cepheid at each wavelength
we plot its magnitude residual from the mean PL relation (as defined
by the Cepheids themselves in each of these samples) as a function of
its measured atmospheric [Fe/H] metallicity. Then for each wavelength
set we solve for the linear sensitivity of that magnitude residual to
Cepheid metallicity. The data and the regression fits are given in
Table 1 and plotted in Figure 2. We also show the residuals for the
reddening-free magnitude W = V- $R_V$(B - V). We note that the
correlation of residuals with metallicity is almost flat at the
H-band, as well as for W.

Viewed in isolation, the scatter at any given wavelength is large and
the slope of any given correlation has a low significance.  However,
the run of slopes as a function of wavelength is clearly
monotonic. This would not reasonably be the case if each of the
measured slopes was statistically insignificant and randomly
distributed around zero.  This raises the possibility that the
relative slopes may not be as poorly determined as is suggested by the
significance of the individual fits alone.

We also note that the scatter is highly correlated across the various
wavelengths. This correlated scatter is not unexpected.  In addition
to any systematic shift of a Cepheid's luminosity due to metallicity
there will be other sources of correlated noise. Intrinsic temperature
(color) differences amongst Cepheids at a fixed period will also
manifest themselves in a similar way. Differential reddening will
scatter Cepheids around the mean (apparent) PL relation.  The
amplitudes of both of these effects are known to be decreasing
functions of wavelength, and although there is no reason to expect
either of them (and certainly not the line-of-sight extinction) to be
functions of the star's metallicity, they will both contribute
significantly to any measured residual deviation of a given star from
the mean PL relation. The back-to-front geometry of the LMC also
contributes correlated scatter. While we cannot unambiguously
disentangle and individually subtract out the effects of reddening,
temperature and geometry, we can take out their averaged and combined
contribution to this sample of stars in order to enhance the
signal-to-noise against which any residual metallicity effect can be
viewed.

In order to re-evaluate the formal error (but not the absolute values)
of the individual slopes at each wavelength we have undertaken to
de-correlate the noise in each of the metallicity-residual plots by
applying the following filter: First, we choose the H-band
data as fiducial. It has essentially zero slope, but still
measureable scatter. We then multiplicatively scale the H-band
residuals and subtract them star-by-star from each of the other
wavelength residual plots, choosing the scale factor that minimizes
the scatter at each particular wavelength. Because metallicity is
independent of period for this sample (see above), and because the
H-band data themselves show no trend with metallicity, there can be no
effect of this scaled subtraction on the slopes of the other
relations. New regressions show this to be the case, but the true
significance of these same slopes now becomes more clear (see Figure 3).

In Figure 4 we plot the sensitivity of the PL relation to metallicity
(i.e., the slopes from Figure 2) as a function of wavelength, but
using the error bars derived from the noise-decorrelated data (which
are given in Table 1). A significant monotonic decline in sensitivity
with metallicity is seen.  The effect is in the sense that at optical
($BV$) wavelengths, there is a positive ($\sim $+0.5 mag/dex)
effect. There are too few Cepheids with $U$-band data to obtain a
statistically significant result, but the overall trend at $U$ is
consistent with a greater sensitivity to metallicity at shorter
wavelengths, albeit with greater uncertainty.  There appears to be
little sensitivity to metallicity at 1-2$\mu$m ($JHK$), and then the
effect reverses sign with $\sim$-0.3 mag/dex at mid-infrared
wavelengths. These results are consistent with those noted by
Romaniello et al. (2008); that is, opposite in sign to most of the
existing empirical calibrations of the metallicity effect.

We do not know the reason for this difference. We note simply that the
current test offers the advantage of high resolution spectroscopic
[Fe/H] abundances for {\it individual} Cepheids, rather than a measure
of the average [O/H] abundance for HII regions at the same azimuthal
distance from the centers of the host galaxies as the Cepheids. It is
also a test based on Cepheids alone, rather than a combination of tip
of the red giant branch and Cepheid distance scales. In any case, this
test provides a completely independent means to place empirical limits
on the sensitivity of the Cepheid Leavitt Law to metallicity.

For the Hubble Space Telescope Key Project (Freedman et al. 2001), a
metallicity correction of -0.2 $\pm$ 0.2 mag/dex was adopted, based on
the empirical results of Kennicutt et al. (1998) and later Sakai et
al. (2004). If instead we adopt the result from this paper, where the
data suggest that the reddening-free W magnitude is not affected by
metallicity then, as noted by Freedman et al. (their Section 8.7) the
Hubble constant is increased by 4\% from 72 to 75~km/sec/Mpc.

\subsection{Deviations from the Hubble Flow}

If metallicity is affecting the magnitudes and colors, and therefore
the derived distances to Cepheids this effect should add (correlated)
scatter into the Hubble diagram. That is, galaxies of higher
metallicity should preferentially scatter in one direction away from
the ridge line in a plot of distance versus velocity when calibrated
by low(er) metallicity Cepheids.  The test is very straight-forward:
as a function of the metallicity of the host galaxy seek out a
correlation of metallicity with deviations in the Hubble diagram,
expressed as deviations in distance modulus.

We have used the reddening-corrected VI Cepheid distance data as
published in Table 3 of the Key Project summary paper (Freedman et
al. 2001, uncorrected for metallicity); and we updated the
flow-corrected velocities using the WEB tool provided by NED (adopting
a Hubble constant of 73~km/s/Mpc), which provides corrections for
perturbations caused by Virgo, the Great Attractor and the Shapley
Supercluster. For the three galaxies in the Virgo Cluster we used a
single velocity appropriate to the cluster as a whole
(957~km/sec). Likewise for the two galaxies in the Fornax Cluster we
use a corrected recession velocity of 1306~km/s.

Figure 5 shows the correlation of deviations in the Hubble diagram
(read off as deviations in distance modulus) plotted as a function of
the HII region [O/H] abundances, measured at the same radial distance
as the Cepheids. There is a considerable amount of residual scatter,
presumably due to random errors in the Cepheid distances combined with
additional peculiar velocities of the parent galaxies over and above
the cluster-induced flows. A formal regression gives the following
solution: $\delta\mu_o$(Cepheid - Hubble Flow) $= -0.17 (\pm 0.31)
([O/H] -8.80) - 0.21 (\pm 0.10$~mag), indicating a mild dependence on
metallicity (with the opposite sign from the dependence found in the
earlier tests in this paper, and opposite in sign to the effect
reported by Romaniello et al. 2008), but with extremely weak
statistical significance. We conclude that the peculiar velocities of
these nearby Cepheid galaxies are sufficiently large, and the sample
of galaxies with Cepheid distances sufficiently small that this test
cannot currently provide a robust test of the metallicity effect.

\section{Discussion and Conclusions}
Spectroscopic [Fe/H] measurements for LMC Cepheids from Romaniello et
al. (2008), combined with multi-wavelength Leavitt relations provide
empirical evidence for a systematic, and wavelength-dependent change
in the luminosities of Cepheids with atmospheric metallicity. The
changes are largest at the bluest wavelengths (amounting to about
+0.5~mag/dex in the B and V bands) and monotonically decrease with
increasing wavelength. The correlation goes flat at about 2$\mu$m
after which the effect reverses in sign and reaches about -0.3~mag/dex
across the mid-IR. This trend is consistent with a purely atmospheric
effect where line-blanketing is known to be greater at blue
wavelengths with the subsequent redistribution of energy resulting in
a slight (energy-balancing) increase in the effective temperature at
longer wavelengths, in this case apparently for wavelengths beyond
2$\mu$m.

This test is distinct from other previous tests for the metallicity
sensitivity of the Leavitt Law to date in that it uses direct
measurements of the Cepheid metallicities on a star-by-star basis,
without recourse to (intermediary) HII region abundances for instance,
or external comparisons (with Population II TRGB distances, for
example). Consistent with the results from Romaniello et al. (2008),
the sign of the effect in this calibration is different from the
aforementioned external tests and calibrations.  R and I-band data for
this set of Cepheids would be of interest in further calibrating this
relationship.

A test of the metallicity sensitivity of the Leavitt Law using
deviations from the pure Hubble flow reveals a very weak signal in
general statistical agreement with other independent tests, but of
very low significance.

Taken on balance we draw the following practical conclusion. Dealing
with metallicity effects on Cepheid distances is best accomplished by
minimizing their impact from the outset by moving the calibrations 
away from the optical to longer wavelengths.  Based on the results
from this test, the cross-over point in the wavelength sensitivity of
the metallicity correction is at near-infrared H or K-band
wavelengths. In any case near- or mid-infrared data are to be
preferred over optical observations for the additional fact that they
significantly reduce the impact of all (foreground Milky Way or
host-galaxy) line of sight extinction. For the immediate future,
moving to the infrared seems to be the best practical solution to a
complicated problem that is  still controversial as to its
magnitude and sign at optical wavelengths.

\medskip
\medskip
\centerline{\it Acknowledgements} This study made use of the NASA/IPAC
Extragalactic Database (NED) which is operated by the Jet Propulsion
Laboratory, California Institute of Technology, under contract with
the National Aeronautics and Space Administration.

\medskip
\medskip

\medskip
\medskip

\noindent
\centerline{\bf References \rm}
\vskip 0.1cm
\vskip 0.1cm

\par\noindent
Bono, G., et al. 2005, \apj, 621, 966

\par\noindent
Bono, G., et al. 2008, \apj, 684, 102

\par\noindent
Efremov, Y.N. 1978, Soviet Astron., 22, 161

\par\noindent
Freedman, W.L., \& Madore, B.F. 1990, ApJ, 365, 186

\par\noindent
Freedman, W.L., {\it et al.} 2001, \apj, 553, 47

\par\noindent
Iben, I., Jr., \& Tuggle, R.S. 1975, \apj, 197, 39

\par\noindent
Kennicutt, R.C., et al. 1998, \apj, 498, 181

\par\noindent
Lee, M.G., Freedman, W.L., \& Madore, B.F. 1993, \apj, 417, 553

\par\noindent
Macri, L.M., Stanek, K.Z., Bersier, D., Greenhill, L.,J., \&  Reid, M.J., 2006, \apj, 652, 1133.

\par\noindent 
Madore, B.F. 1985, ``Cepheids: Theory and Observations'', IAU Coll. No. 82, Cambridge Univ Press, pg. 166

\par\noindent
Madore, B.F., et al., 2009, \apj, 695, 988.

\par\noindent 
Magnier, E.A., Prins, S., Augusteijn, T., van Paradijs,
J., \& Lewin, W.H.G. 1997, \aap, 336, 442

\par\noindent 
Persson, S.E., Madore, B.F., Krzeminski, W., Freedman, W.L., Roth, M.,
\& D. C. Murphy, D.C. 2004, /aj, 128, 2239 

\par\noindent 
Romaniello, M., et al. 2008, \aap, 488, 731


\par\noindent
Robertson, J.W. 1973 \apj, 185, 817

\par\noindent
Sakai, S., et al. 2004, \apj, 608, 42

\par\noindent
Scowcroft, V., Bersier, D., Mould, J.R., \& Wood, P.R.  2009, \mnras, 396, 1287 

\par\noindent
Welch, D.L., McLaren, R.A., Madore, B.F., \& McAlary, C.W. 1987, \apj, 321, 162

\begin{deluxetable}{cccc}




\tablecaption{Wavelength Dependence of Metallicity Sensitivity }

\tablenum{1}


\tablehead{\colhead{Bandpass} & \colhead{slope} & \colhead{$\sigma$} & \colhead{$R^2$} \\ 
\colhead{} & \colhead{(mag/dex)} & \colhead{(mag/dex)} & \colhead{}} 

\startdata
B & +0.59 & $\pm$0.49 & 0.37\\
V & +0.50 & $\pm$0.31 & 0.27\\
J & +0.14 & $\pm$0.07 & 0.12\\
H & +0.05 & $\pm$0.02 & fiducial\\
K & +0.02 & $\pm$0.03 & 0.10\\
3.6 & $-$0.39 & $\pm$0.16 & 0.77\\
4.5 & $-$0.25 & $\pm$0.18 & 0.77\\
5.8 & $-$0.39 & $\pm$0.17 & 0.85\\
8.0 & $-$0.38 & $\pm$0.16 & 0.88\\
\enddata




\end{deluxetable}

\vfill\eject
\end{document}